\documentclass[superscriptaddress,nofootinbib,twocolumn,prb]{revtex4}

\usepackage{amsmath,amssymb}
\usepackage{graphicx}
\usepackage{color}
\usepackage{hyperref}
\usepackage{url}
\usepackage[utf8]{inputenc}
\usepackage{slashed}
\usepackage{subfigure}
\usepackage{slashed,bbm}
\usepackage{graphics,psfrag,epsfig}
\usepackage{dsfont}
\usepackage{setspace}

\begin{document}

\title{Gauge-field-assisted Kekul\'e quantum criticality}

\author{Michael~M.~Scherer}
\affiliation{Department of Physics, Simon Fraser University, Burnaby, Canada}
\affiliation{Institut f\"ur Theoretische Physik, Universit\"at Heidelberg, Philosophenweg 16, 69120 Heidelberg, Germany}

\author{Igor~F.~Herbut}
\affiliation{Department of Physics, Simon Fraser University, Burnaby, Canada}

\begin{abstract}
We study the quantum phase transition of $U(1)$ - charged Dirac fermions Yukawa-coupled to the Kekul\'e valence bond solid order parameter with $Z_3$ symmetry of the honeycomb lattice. The symmetry allows for the presence of the term in the action which is cubic in the Kekul\'e order parameter, and which is expected to prevent the quantum phase transition in question from being continuous. The Gross-Neveu-Yukawa theory for the transition is investigated using a perturbative renormalization group and within the $\epsilon$ expansion close to four space-time dimensions. For a vanishing $U(1)$ charge we show that quantum fluctuations may render the phase transition continuous only sufficiently far away from 3+1 dimensions, where the validity of the conclusions based on the leading order $\epsilon$~expansion appear questionable. In the presence of a fluctuating gauge field, on the other hand, we find quantum critical behavior even at weak coupling to appear close to 3+1 dimensions, that is, within the domain of validity of the perturbation theory. We also determine the renormalization group scaling of the cubic coupling at higher-loop orders and for a large number of Dirac fermions for vanishing charge.
\end{abstract}

\maketitle

\section{Introduction}


The research on Dirac materials~\cite{vafek2013, wehling2014} encompasses many areas of physics, ranging from a fundamental theory with ties to particle physics, to concrete sought-after technological applications in materials science. From a theoretical point of view, interacting
Dirac systems are particularly appealing as a playground for development of novel and applications of known  quantum-field-theoretical tools, providing an opportunity for their testing against numerical and experimental results.

\begin{figure}[t!]
\includegraphics[height=0.45\columnwidth]{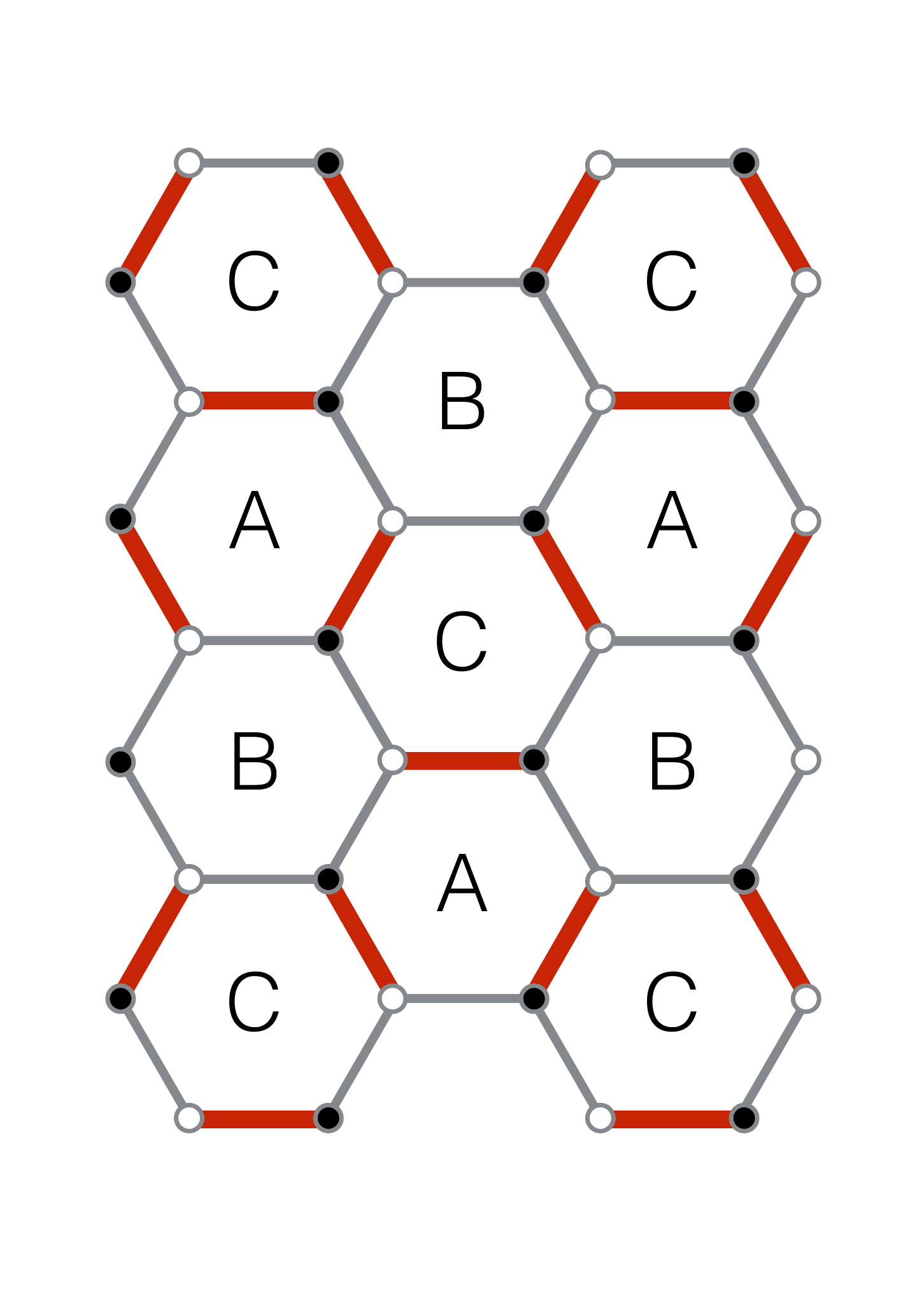}
\hspace{0.2cm}
\includegraphics[height=0.45\columnwidth]{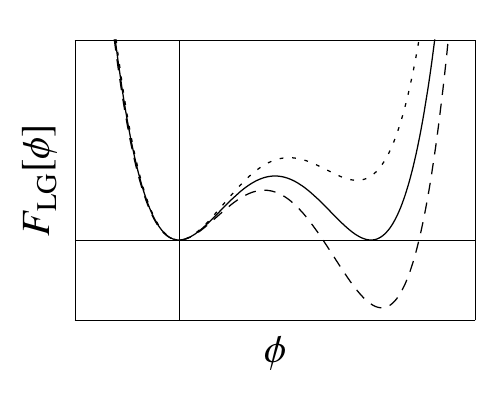}
\caption{Left panel: An example of the Kekul\'e dimerization pattern. Shorter bonds are marked by thicker red lines and longer bonds by the thinner gray lines. Right panel: LG free energy across the phase transition, where the dotted (dashed) line corresponds to the disordered (ordered) phase, with $r>r_c\ (r<r_c)$.}
\label{fig:kekule}
\end{figure}

Graphene\cite{Novoselov2005,Geim2007,castroneto2009}, in particular, at charge-neutrality serves as a prime example of a class of materials where Dirac nature of the low-energy excitations emerges naturally from the underlying honeycomb lattice structure, and is further protected by its symmetries.
The presence of sufficiently strong interactions can lead to spontaneous symmetry breaking signaled by the appearance of an ordered ground state.
There is a wide variety of possible ordering patterns determined by small variations of the short-ranged interaction parameters, which have been analyzed by different theoretical approaches, ranging from mean-field calculations to the renormalization group and numerical simulations\cite{sorella1992,herbut2006,honerkamp2008,herbut2009, raghu2008, juricic2009, assaad2013, chandra2013, parisen2014, janssen2015, sorella2015, wang2016, hasselmann2016, volpez2016,sanchez2016}.
A large repulsive onsite interaction $U$, for example, exhibits a continuous quantum phase transition from the semimetallic (SM) phase into an antiferromagnetic spin-density-wave (SDW) state, whereas a nearest-neighbor interaction $V_1$ would induce a conventional charge-density wave (CDW) state.
While there is an ongoing debate on whether a pure second-nearest-neighbor interaction $V_2$ can induce a topological Mott insulating state\cite{raghu2008,duric2014} or whether it favors a sublattice charge-modulated state\cite{grushin2013,daghofer2014, capponi2015, motruk2015}, 
it is believed that the interplay between balanced $V_1$ and $V_2$ gives rise to yet another ordered ground state -- the Kekul\'e valence bond solid (VBS)~\cite{hou2007, roy2010c, weeks2010, garciamartinez2013}. Alternatively, this ordering can also emerge from electron-phonon interactions\cite{nomura2009,kharitonov2012,Classen2014b}.
The translation symmetry breaking Kekul\'e VBS corresponds to a particular dimerization pattern of the fermions on the honeycomb lattice with a bond-dependent nearest-neighbor hopping amplitude (see Fig.~\ref{fig:kekule}), and has recently been accessed even experimentally~\cite{Gutierrez2016}.


The Kekul\'e ordering pattern can be described by a $Z_3$ symmetric order parameter that describes breaking of the lattice translational symmetry and the $C_3$ rotational symmetry\cite{li2015}. Due to $Z_3$ symmetry the cubic terms in the order parameter in the Landau-Ginzburg (LG) free energy are allowed, which generically prevent a phase transition toward the ordered state from being continuous\cite{Golner1973}.
The LG free energy $F_{\text{LG}}$ takes the form
\begin{align}\label{eq:LGfe}
	F_{\text{LG}}[\phi]=F_0+r\, |\phi|^2 + g (\phi^3+\phi^{\ast 3})+\lambda |\phi|^4\,,
\end{align}
where $\phi$ represents the complex-valued order parameter of the VBS, $F_{\text{LG}}$ is symmetric under the field transformation $\phi \to \exp(2\pi i\, n/3)\phi, n\in \mathbbm Z$ and the parameter $r$ is used to tune through the phase transition by crossing a critical value~$r=r_c$.
In the case when the cubic coupling $g$ vanishes, the LG theory suggests that the phase transition is of second order, since at $\phi=\phi_0\neq0$ a non-vanishing minimum of the LG free energy emerges continuously from the origin at $\phi=0$.
For a finite~$g$, on the other hand, the minimum of $F_{\text{LG}}$ jumps discontinuously from $\phi=0$ to a $\phi_0\neq0$ as $r$ goes through its critical value~$r_c >0$, i.e. a first-order transition is expected (see Fig.~\ref{fig:kekule}). This simple consideration is not sufficient, however, to describe the Kekul\'e phase transition when massless fermions are present and coupled to the $Z_3$ order parameter.


We study the quantum phase transition of massless Dirac fermions toward the Kekul\'e VBS within the appropriate Gross-Neveu-Yukawa theory, and in $D=3+1-\epsilon$ dimensions. We find that the presence of (neutral) Dirac fermions tends to reduce the first-order character of the phase transition\cite{li2015}, and that closer to $D=2+1$ dimensions the phase transition may even be rendered continuous\cite{Roy2013c}. However, the values of parameter $\epsilon$ where this happens are not small, and this possibility seems beyond the realm of control of the $\epsilon$ expansion. Sizable corrections which may even overturn the conclusion based on the lowest order calculation
can generally be expected beyond the one-loop order. Adding a $U(1)$ gauge field that would couple exclusively to fermions, as the Kekul\'e order parameter is electrically neutral, on the other hand, significantly further reduces the discontinuous nature of the transition, and we find that in its presence the transition may become continuous even at small values of $\epsilon$. The resulting quantum critical behavior in this case is discussed in some detail.
We also present an argument that in the limit of a large number of fermion species the phase transition becomes continuous only for $\epsilon >1$,
even if higher-loop terms are included.


The paper is organized as follows:
In Sec.~\ref{sec:model}, we introduce a continuum model for Dirac fermions coupled to a $U(1)$ gauge field.
We also discuss the parametrization of the Kekul\'e order parameter and its coupling to the fermionic sector.
In Sec.~\ref{sec:rge}, the renormalization group equations for the full model are presented and their fixed-point solutions are studied in Sec.~\ref{sec:FPana}.
Conclusions are drawn in Sec.~\ref{sec:conc}.

\section{Model}\label{sec:model}


The motion of non-interacting fermions on the honeycomb lattice can be most simply described in terms of a tight-binding Hamiltonian
\begin{align}\label{eq:h0}
H_0=-t\sum_{\vec{R},i}\left[u^\dagger(\vec{R}) v(\vec{R}+\vec{\delta}_i)+\text{h.c.}\right]\,,
\end{align}
with nearest-neighbor hopping amplitude $t$.
Here, $u$ and $v$ are the electron annihilation operators
at the two triangular sublattices of the lattice and the sum runs over the sites $\vec{R}$ of the first triangular sublattice with position vectors $\vec{R}_1^T=a\left(\sqrt{3}/2,-1/2\right)$ and $\vec{R}_2^T=a\left(0,1\right)$.
The lattice spacing $a$ is set to $a=1$ and the three nearest-neighbor vectors $\vec{\delta}_i$, located on the second sublattice, explicitly read $\vec{\delta}_1^T=\left(1/(2\sqrt{3}),1/2\right)$, $\vec{\delta}_2^T=\left(1/(2\sqrt{3}),-1/2\right)$ and $\vec{\delta}_3^T=\left(-1/\sqrt{3},0\right)$.
Diagonalization of the free Hamiltonian $H_0$ provides a spectrum with two energy bands and band dispersion $\epsilon_{\vec{k}}=\pm |\sum_{i=1}^3\exp(\vec{k}\cdot\vec{\delta}_i)|$.
At the corners of the Brillouin zone, given by the two points $\vec{K}^T=(2\pi/\sqrt{3},2\pi/3)$ and $\vec{K}^\prime=-\vec{K}$, the two energy bands touch linearly and isotropically. Here, the energy dispersion gives rise to two inequivalent Dirac points.


Retaining only the Fourier modes near the band touching points a continuum low-energy effective theory for $H_0$ can be written down in terms of a free Dirac Lagrangian\cite{Semenoff1984}, $\mathcal{L}_{\psi,0}=\bar\psi\,\gamma_\mu\partial_\mu\psi$,
where $\partial_\mu=(\partial_\tau, \vec{\nabla})$ and the $\gamma$~matrices are defined in a four-dimensional representation as $\gamma_0=\mathbbm{1}\otimes\sigma_z$, $\gamma_1=\sigma_z\otimes\sigma_y$, and $\gamma_2=\mathbbm{1}\otimes\sigma_x$ in $D=2+1$ dimensions.
The $\sigma_i$ are the conventional Pauli matrices and the conjugate of the Dirac field is given by $\bar\psi=\psi^\dagger\gamma_0$. We consider the Dirac field in Fourier space $\psi(x)=\int d^Dqe^{iqx}\psi(q)$ and express it in terms of Grassmann fields $u,v$ by\cite{herbut2006}
$\psi^\dagger(q)=\left[u^\dagger(K+q),v^\dagger(K+q),u^\dagger(-K+q),v^\dagger(-K+q)\right]$ with momentum vectors $q=(\omega,\vec{q})$ gathering Matsubara frequency $\omega$ and wavevector $\vec{q}$.
The reference frame is chosen such that $q_x=\vec{q}\cdot \vec{K}/|\vec{K}|$ and $q_y=(\vec{K}\times\vec{q})\times \vec{K}/|\vec{K}|^2$.
We define two more $\gamma$ matrices which anticommute with all $\gamma_\mu$, reading $\gamma_3=\sigma_x\otimes\sigma_y$ and $\gamma_5=\sigma_y\otimes\sigma_y$. We define $\gamma_{35}=-i\gamma_3\gamma_5$, which commutes with all $\gamma_\mu$ and anticommutes with $\gamma_3$ and $\gamma_5$.
As a convenient generalization of this model, we introduce an arbitrary number of pairs of Dirac points $\pm\vec{K}_i$ into the spectrum with $i=1,...,N$ and refer to $N$ as the number of fermion flavors of four-component spinors.
We note that the case of spin-1/2 fermions on the honeycomb lattice with spin-rotation invariance as, e.g., relevant for electrons in graphene, can be described by setting the number of fermion flavors to $N=2$ corresponding to an eight-component spinor.


We also introduce a coupling of the fermions to an electromagnetic $U(1)$ gauge field $A_\mu$.
This is formally done by minimal coupling, leading to the Lagrangian\cite{Roy2013c}
\begin{align}
	\mathcal{L}_\psi&=\bar\psi\,\gamma_\mu\left(\partial_\mu-ieA_\mu\right)\psi\,,
\end{align}
with the $U(1)$ charge coupling $e$, which describes the interaction of fermions with the electromagnetic field in terms of a local quantum field theory.

The kinetic term of the gauge field reads
\begin{align}
\mathcal{L}_{\text{em}}=\frac{1}{4}F_{\mu\nu}F_{\mu\nu}\,,
\end{align}
with $F_{\mu\nu}=\partial_\mu A_\nu-\partial_\nu A_\mu$ and we further add a gauge fixing term to the Lagrangian $\mathcal{L}_{\text{gf}}=\frac{1}{2\xi}(\partial_\mu A_\mu)^2$ with gauge fixing parameter $\xi$. Here, $\xi=0$ corresponds to Landau gauge and $\xi=1$ corresponds to Feynman gauge.
In the following, we will discuss two distinct cases\cite{gonzalez1993,Gorbar2001,Herbut2001,Herbut2013}: (1) The electromagnetic field propagates in three-dimensional space, i.e. in 3+1 dimensions whereas the propagation of the fermions is restricted to a lower dimension, e.g., $D=2+1$. (2) The electromagnetic field is defined such that it propagates in the same spacetime dimensions as the fermions, i.e. $D=d+1$.

\subsection{Kekul\'e valence bond solid}


The Kekul\'e VBS corresponds to a particular dimerization pattern of the fermions on the honeycomb lattice that can be captured by adding a bond-dependent modification of the hopping strength~\cite{hou2007} to the simple tight-binding Hamiltonian
\begin{align}\label{eq:Deltah0}
\Delta H_0=-\sum_{\vec{R},i}\left[\Delta t_{\vec{R},i}u^\dagger(\vec{R}) v(\vec{R}+\vec{\delta}_i)+\text{h.c.}\right]
\end{align}
where $\Delta t_{\vec{R},i}=\phi(\vec{R}) e^{i \vec{K}\cdot\vec{\delta}_i}e^{i (\vec{K}-\vec{K}^\prime)\cdot\vec{R}}/3+\text{c.c.}$ and $\phi$ is a complex-valued order parameter possessing $Z_3$ symmetry.
This modified hopping introduces a chiral mixing between the two Dirac points and gives rise to a phase transition from a Dirac semimetal towards a Kekul\'e VBS with a single-particle mass gap.
In terms of the low-energy effective field theory, the coupling between the fermions and the order parameter field reads\cite{roy2010c}
\begin{align}\label{eq:lagb}
	\mathcal{L}_y=\,y\left(\phi_1\bar\psi\, i\gamma_3\psi+\phi_2\bar\psi\, i\gamma_5\psi\right)\,,
\end{align}
with $\phi=\frac{1}{\sqrt{2}}(\phi_1+i\phi_2)$ and $\phi_1,\phi_2$ are real fields.
The dynamics of the bosonic order parameter is described by the Lagrangian
\begin{align}\label{eq:lagb}
	\mathcal{L}_\phi=\phi^\ast\left(\partial_\tau^2-\nabla^2+ m^2\right)\phi+ g\left(\phi^3+\phi^{\ast 3}\right)+\lambda|\phi|^4\,
\end{align}
which due to $Z_3$ symmetry allows the cubic term $\propto g$.
We note that the Kekul\'e order parameter is gauge neutral, and therefore its Lagrangian is not modified by the presence of the $U(1)$ gauge field.
This is in contrast to a superconducting order parameter, which combines two electrons into a Cooper pair and therefore comes with a charge\cite{Roy2013c,Zerf2016} of $2e$.


In sum, we study a $U(1)$ gauge theory for $N$ flavors of Dirac fermions in $D=d+1$ dimensions coupled to a charge-neutral complex order parameter corresponding to a Kekul\'e distortion.
The full Lagrangian $\mathcal{L}$ is composed of a free fermionic part $\mathcal{L}_\psi$, a bosonic part $\mathcal{L}_\phi$, a Yukawa interaction $\mathcal{L}_y$, the Maxwell Lagrangian $\mathcal{L}_{\text{em}}$ and the gauge fixing $\mathcal{L}_{\text{gf}}$, i.e. the action is given by
\begin{align}\label{eq:action}
S=\int d\tau d\vec{x}^{D-1}\left[\mathcal{L}_\psi+\mathcal{L}_\phi+\mathcal{L}_y+\mathcal{L}_{\text{em}}+\mathcal{L}_{\text{gf}}\right]\,.
\end{align}

\section{Renormalization group equations}\label{sec:rge}

A continuous phase transition and the concomitant (quantum) critical behavior is related to the presence of an infrared attractive fixed point in the system's renormalization group equations.
At a RG fixed point the system becomes scale invariant and the free energy exhibits a scaling form giving rise to universal critical exponents\cite{herbutbook}.
To approach the critical point of a phase transition one parameter has to be tuned to its critical value, e.g., the temperature $T\to T_c$ or, in the case of a quantum critical point, the doping level or a coupling strength.
In the RG framework a tuning parameter corresponds to an RG relevant direction, which -- unless precisely tuned -- drives the system away from its fixed point.

Canonical power counting determines the RG relevance of the parameters of a model close to its non-interacting limit, i.e. near the Gaussian fixed point.
At an non-Gaussian fixed point, however, the canonical power counting receives corrections from the interactions which can affect the relevance of the RG directions.
A positive (negative) power counting dimension suggests a relevant (irrelevant) parameter.
Parameters with vanishing power counting dimension are called marginal.
For instance the power counting dimension of the mass parameter in the Lagrangian $\mathcal{L}_\phi$ is $[m^2]=2$, i.e. it is strongly relevant and therefore needs to be fine-tuned to approach the critical point.
The power counting dimension of the quartic coupling $\lambda$ is $[\lambda]=\epsilon$ in $D=4-\epsilon$ dimensions.
It is only slightly positive and is turned negative at a non-Gaussian fixed point even if the interactions are perturbatively small and therefore this coupling automatically approaches its fixed point value toward the infrared.
The same reasoning also holds for the Yukawa coupling $y$ and the electromagnetic coupling $e$.
In most condensed matter models of similar type, this exhausts the available RG parameters leaving only the squared mass as an RG relevant direction, i.e. a single tuning parameter to approach the quantum critical point.

Here, however, the cubic order parameter terms introduce another RG relevant direction with a large power-counting dimension $[g]=1+\epsilon/2$.
In case only one tuning parameter is available, no scaling behavior, as induced by the vicinity to a RG fixed point, will be observed.
On the other hand, the quantum critical behavior of low-dimensional Dirac fermions coupled to order parameters is governed by non-trivial fixed points where the canonical power counting is modified.
When these modifications are large enough, they can flip the sign of the canonically relevant coupling $g$ and therefore restore the continuous transition.
We investigate this scenario and analyze its domain of applicability and control.

For  the  above  action $S$ and  at  zero  temperature we calculate the renormalization group (RG) equations in the Wilsonian scheme by simultaneously integrating out the fermionic as well as the bosonic modes within the narrow momentum shell $\Lambda/b< (\omega^2+\vec{q}^2)<\Lambda$.
At one-loop order in $D=4-\epsilon$ dimensions RG equations for the squared cubic and the quartic bosonic couplings appearing in $\mathcal{L}_\phi$, at $m=0$ read
\begin{align}\label{eq:beta01}
	\beta_{g^2}&=\frac{d g^2}{d \ln b}=(2+\epsilon-3Ny^2-6\lambda-\frac{27}{2}g^2)g^2\,,\\
	\beta_\lambda&=\epsilon\lambda-2Ny^2(\lambda-y^2)-5\lambda^2-162g^4+63\lambda g^2\,.\label{eq:beta02}
\end{align}
Here, we have rescaled the couplings $g^2\to g^2/(4\pi^2 \Lambda^{2+\epsilon})$ and $\lambda\to \lambda/(4\pi^2 \Lambda^{\epsilon})$.
The one-loop $\beta$ functions of the purely bosonic couplings are independent of the charge $e$ as it does not directly couple to the neutral boson field.
The squared Yukawa coupling $y^2$ flows as
\begin{align}\label{eq:betay}
	\beta_{y^2}&=\frac{d y^2}{d \ln b}=\epsilon y^2-(N+1)y^4-\frac{9}{2}g^2y^2+6e^2y^2\,,
\end{align}
where we have rescaled $y^2\to y^2/(4\pi^2 \Lambda^{\epsilon})$.
Here, we obtain contributions to the $\beta$ function from the $U(1)$ field due to a non-trivial renormalization of the Yukawa vertex, as reflected in the last term in Eq.~\eqref{eq:betay}.

Finally, we consider the $\beta$ function of the charge parameter for the case of a gauge field propagating in $3+1$ dimensions whereas fermions and complex bosons propagate on a lower-dimensional brane with $(3-\epsilon) +1$ dimensions~\cite{gonzalez1993,Gorbar2001,Herbut2001,Herbut2013}.
In this case, the bare gauge field propagator of the effective $(3-\epsilon) +1$-dimensional theory is obtained from ``integrating out" the additional spatial direction, which is most conveniently performed in Feynman gauge which leaves the effective gauge parameter invariant when going from 3+1 to $(3-\epsilon) +1$  dimensions. As a result one obtains a bare Maxwell term which is non-analytic in momentum $\propto |q|^{2-\epsilon}$, instead of the conventional, analytic,  $q^2$~dependence\cite{Gorbar2001}. Due to the resulting non-analyticity of the effective gauge field propagator at $q=0$ the charge coupling does not receive renormalization corrections and therefore becomes an exactly marginal coupling;  $\beta_{e^2}=0$. The $\beta$ function for $y^2$, on the other hand, remains unaffected by the above consideration.

We will also consider the $\beta$ function of the charge for the case that the $U(1)$ gauge field propagates in the same dimension as the fermions. In $D=3+1-\epsilon$ and with $e^2\to e^2/(4\pi^2 \Lambda^{\epsilon})$ it is then given by
\begin{align}
	\beta_{e^2}=\frac{d e^2}{d \ln b}=\epsilon e^2-\frac{4}{3}N e^4\,.\label{eq:betae}
\end{align}
%

\section{Fixed point analysis}\label{sec:FPana}

The set of the four $\beta$ functions, i.e. $\beta_{e^2},\beta_{\lambda},\beta_{y^2}$ and $\beta_{g^2}$, features various fixed points, e.g., the Gaussian (G) fixed point with $e^{2}_\ast =y^{2}_\ast=g^{2}_\ast=\lambda_\ast=0$ and the purely bosonic Wilson-Fisher (WF) fixed point with  $e^{2}_\ast =y^{2}_\ast=g^2_\ast =0\,,\ \lambda_\ast=\epsilon/5$.
In particular, we note that the $\beta$ function for the cubic coupling, Eq.~\eqref{eq:beta01}, always has the trivial fixed-point solution, $g^{2}_\ast=0$, even if the other couplings have non-vanishing fixed-point values.
In case $g^2=0$, a finite value for $g^2$ cannot be generated by the RG flow as the $Z_3$ symmetry of the model is replaced by a larger $U(1)$ symmetry, which is preserved under RG transformations. Here, we will discuss exclusively the Gross-Neveu-Yukawa fixed-points with $g^{2}_\ast=0$ that govern the quantum critical behavior of the model.

\subsection{Exactly marginal charge coupling}

We consider first the scenario where the charge is exactly marginal, and therefore can be taken as a fixed parameter of the model.
The fixed-point equation for the Yukawa coupling then reads
\begin{align}
y^{2}_\ast=\frac{\epsilon}{N+1}+\frac{6\bar e^2}{N+1}
\end{align}
where we have split the expression into a term without the gauge-field coupling $\propto \epsilon$ and a term exclusively originating from a non-vanishing charge.
Accordingly, Eq.~\eqref{eq:beta02} for the four-boson coupling yields a fixed-point value reading
\begin{align}
\lambda_{\ast}=\frac{1-N+R_{N}}{10(N+1)}\epsilon+\frac{6(S_{N}-N)}{5(N+1)}\bar e^2+\mathcal{O}(\epsilon^2,\bar e^4)\,,
\end{align}
to first order in $\epsilon$ with $R_{N}=\sqrt{1+38N+N^2}$ and $S_{N}=\sqrt{N^2+10N}$.
Here, we have also expanded around vanishing $\bar e^2$ as we consider the charge to be a small perturbation.
We note that the limits $\bar e^2\to 0$ and $\epsilon \to 0$ have to be taken with some care, as they do not commute.
We have chosen the order of limits in such a way as to recover best the behavior for $\epsilon \in [0,1]$ and small $\bar e^2 \lesssim 0.12$.
In the limit $\bar e^2 \to0$ our results for $y^2_\ast$ and $\lambda_\ast$ exactly recover the fixed-point values of the neutral model.
The RG scaling related to the interaction parameters $y^2$ and $\lambda$ can be derived from diagonalization of the corresponding block of the stability matrix
\begin{align}\label{eq:stabmat}
	\mathcal{M}=
	\begin{pmatrix}
	\frac{\partial\beta_{y^2}}{\partial y^2} & \frac{\partial\beta_{\lambda}}{\partial y^2}\\
	\frac{\partial\beta_{y^2}}{\partial \lambda} & \frac{\partial\beta_{\lambda}}{\partial \lambda}
	\end{pmatrix}\bigg|_{\bar e^2,y^2_\ast,\lambda_\ast,g_\ast^2}\,,
\end{align}
and we find the two negative eigenvalues
\begin{align}
m_1&=-\epsilon-6 \bar e^2\,,\quad m_2= -\frac{R_N}{N+1}\epsilon-\frac{12S_N}{N+1}\bar e^2 \,.
\end{align}

The RG scaling of the squared mass of the order parameter constitutes the first RG relevant direction of the model and determines the inverse correlation length exponent at the fermionic fixed point.
It is given by the relation
\begin{align}\label{eq:num1}
\theta_{m^2}=\nu^{-1}=2-Ny^{2}_\ast -2\lambda_{\ast}\,.
\end{align}
As the Yukawa coupling can become large for sizable charge this can reduce considerably the value of the RG scaling of the squared mass as compared to its canonical value of 2. We note that $\theta_{m^2}$ remains positive for $\bar e^2 \lesssim 0.12$ for all values of $N$ and $D\in [3,4]$.
Therefore, the bosonic mass parameter serves as a relevant RG direction providing the tuning parameter for the quantum phase transition.

\subsubsection{RG scaling of the cubic coupling}

The $\beta$ function of $g^2$ is proportional to $g^2$ itself
and so the RG scaling due to the cubic term at $g_\ast =0$ is given directly by the derivative of $\beta_{g^2}$ with respect to $g^2$ evaluated at the fermionic fixed point
\begin{align}\label{eq:theta2e2b}
	\theta=\frac{\partial \beta_{g^2}}{\partial g^2}\Big|_{\bar e^2,y^2_\ast,\lambda_\ast,g_\ast^2}\,.
\end{align}
The phase transition to the Kekul\'e VBS state can be rendered continuous when $\theta$ becomes negative. We calculate $\theta=\theta(\bar e^2)$ to order $\epsilon$,
\begin{align}
 \theta&=2+(1-3A_N)\epsilon-18\bar e^2  B_N\label{eq:theta2bare}\,,
\end{align}
where we have defined
\begin{align}
A_N&=\frac{1+4 N+R_N}{5 (N+1)}\,,\quad B_N=\frac{3N+2S_N}{5(N+1)}\,.
\end{align}
The function $A_N$ is positive, and in particular, $A_N\geq 1$ for $N\geq1/2$. To the leading order, the expression in Eq.~\eqref{eq:theta2bare} may become negative then with the increase of $\epsilon$, allowing the transition to become continuous.

\begin{figure}[t!]
\includegraphics[width=1.0\columnwidth]{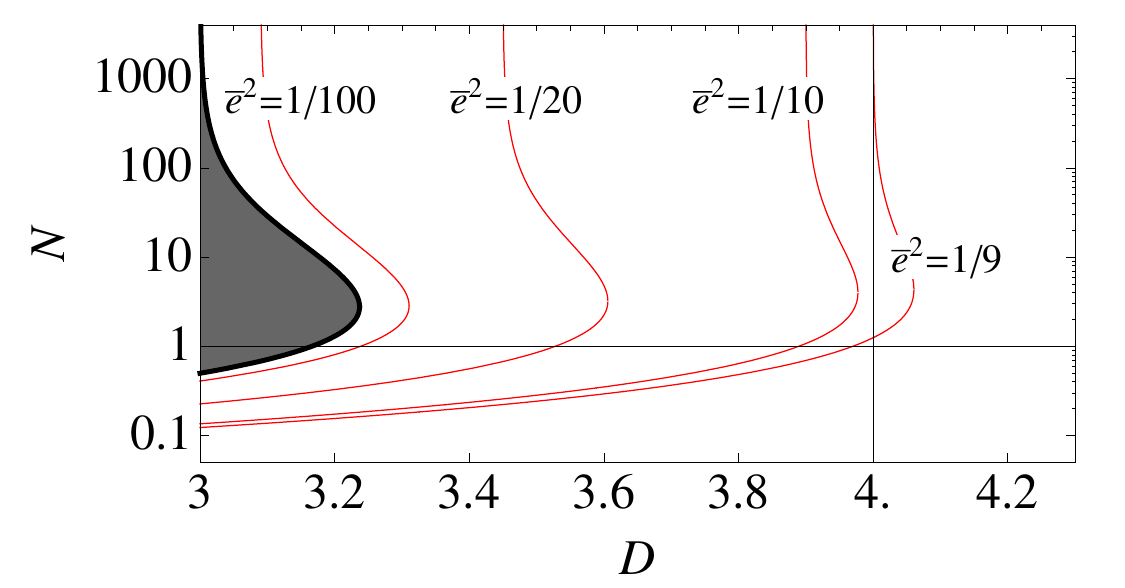}
\caption{Sign of the dimension of the cubic coupling as a function of the dimensionality $D$ and the number of fermion flavors $N$ in first-order $\epsilon$ expansion with a fixed value of the charge $e^2$. The gray-shaded area shows the region in $(D,N)$ space where $\theta(\bar e^2=0)$ is negative, and the cubic term irrelevant. Quite a large values for $\epsilon$ are evidently needed in this case.
The behavior for finite values of $\bar e^2 \in \{1/100,1/20,1/10,1/9\}$ is indicated by the red lines. To the left of these lines, we have $\theta<0$ and the Kekul\'e transition being continuous. Finite charge $\bar e^2$ substantially decreases the required value of $\epsilon$ that would render $\theta<0$.}
\label{fig:theta2e}
\end{figure}

The case $\bar e^2=0$, deserves special consideration: For $N<1/2$, we find $A_N<1$ and therefore $\theta(\bar e^2=0)$ is positive for any $\epsilon \leq 1$.
On the other hand, at $N=1/2$, we have $\theta(\bar e^2=0)=2-2\epsilon$, rendering $\theta=0$  for $\epsilon=1$.
Furthermore, at $\epsilon=1$, we find that for $N\geq 1/2$ the exponent $\theta$ is negative.
In the limit of large $N$, on the other hand, $\theta(e^2=0,N\to \infty)=2-2\epsilon$.
Fig.~\ref{fig:theta2e} shows that a negative $\theta(e^2=0)$ generally requires quite a large value of $\epsilon$ for all values of $N$.
For such large values of $\epsilon$ obviously it becomes questionable whether the conclusion about the relevance of the cubic coupling based on the leading order calculation is trustworthy.

For a finite charge one needs to include the function $B(N)$,  which is always positive, with $B_N\to 0$ for $N\to 0$. We find $B_{N=1}\approx 0.96$, $B_{N=2}\approx 1.05$ and $B_{N\to\infty}=1$, for example. A finite charge $\bar e^2\neq 0$ therefore only helps the tendency of rendering the cubic term irrelevant by gapless fermions. This is clearly exhibited in Fig.~\ref{fig:theta2e} for several specific values of $\bar e^2$, where one sees how the quantum critical region can be brought considerably closer to the perturbatively accessible regime of small~$\epsilon$.
A special limit is given in the case $N\to \infty$, where $A_{N\to\infty}=B_{N\to\infty}=1$. Here, for the choice $\bar e^2=1/9$, we have $\theta=-2\epsilon$ and $\theta_{m^2}=4/3-\epsilon$.

\subsection{Two-loop analysis for $N\to\infty$ for $\bar e^2=0$}

At two-loop order and at $\bar e^2=0$, the $\beta$ function for the cubic coupling squared  is composed from its one-loop and two-loop contributions, $\beta_{g^2}=\beta_{g^2}^{\text{(1-loop)}}+ \beta_{g^2}^{\text{(2-loop)}}$.
In the large-$N$ limit, we can deduce the exponent $\theta$ at two-loop order without accessing the full set of $\beta$ functions with all numerical coefficients at that order.
Recall that the fixed point value of $g_{\ast}^2=0$ is protected by symmetry to arbitrary order.
Due to the vertex structure of the model, all diagrams contributing to $\beta_{g^2}$ at two-loop order are exactly cubic in the couplings $y^2,g^2, \lambda$, e.g., they are porportional to $(g^2)^3, (g^2)^2 \lambda$, $(g^2) (y^2) \lambda,...\,$.
However, no diagrams that are of order $(g^2)^0$ appear, as these would generate a finite $g^2$ from $g^2=0$.

\begin{figure}[b!]
\includegraphics[width=.7\columnwidth]{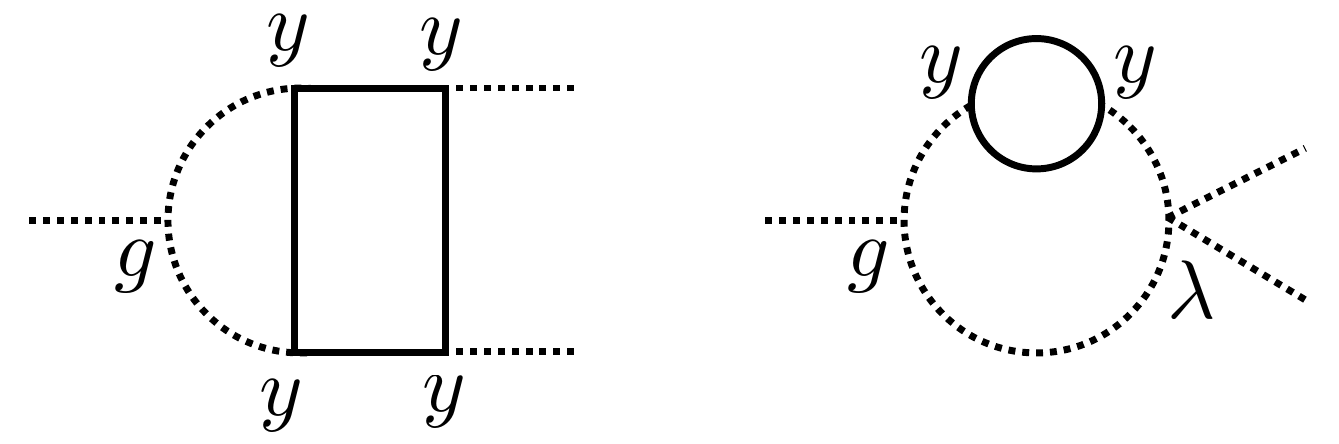}
\caption{Diagrams contributing to the two-loop $\beta$ function of the cubic coupling. Here, the dashed lines represent bosons and the solid lines are fermions. The closed fermion loops come with a factor $N$. We do not display two-loop diagrams without $N$ multiplicity.}
\label{fig:2loop}
\end{figure}

Diagrams that are of order $\mathcal{O}\left((g^2)^2\right)$ and higher cannot contribute to $\theta$, cf. Eq.~\eqref{eq:theta2e2b}, as their contributions vanish at the fixed point where $g_\ast^2=0$.
Therefore, the only two-loop diagrams in $\beta_{g^2}$ from the class of cubic terms in $y^2,g^2, \lambda$ that can contribute to $\theta$ are the ones that are linear in $g^2$, i.e.,
\begin{align}\label{eq:fast2l}
\hspace{-0.18cm}\Delta\beta_{g^2}^{\text{(2-loop)}}=g^2 \left(a\lambda^2+(b +c N) \lambda y^2+ (d+eN) y^4\right),
\end{align}
with $\{a,b,c,d,e\} \in \mathbbm{R}$.
Here, we have already introduced the appropriate factors of $N$ that can be deduced from simple diagrammatic considerations, i.e. appearing closed fermion loops in the two-loop diagrams, cf.~Fig.~\ref{fig:2loop}.

The fixed point coordinates of the couplings $\lambda, y^2$ to two-loop order are identical to the ones of the uncharged chiral XY model without the cubic term~\cite{Rosenstein:1993zf}.
The large-$N$ behavior of the fixed point coordinates can be inferred from a canonical consideration of the two-loop contributions to the $\beta$ functions of  $\lambda$ and $y^2$,
\begin{align}
	\Delta\beta_{y^2}^{\text{(2-loop)}}&= a_y\lambda^2 y^2+b_y \lambda y^4+(c_y+N d_y)y^6\,,\\
	\Delta\beta_{\lambda}^{\text{(2-loop)}}&= a_\lambda\lambda^3 +N b_\lambda\lambda^2 y^2+N c_\lambda \lambda y^4+N d_\lambda y^6,\nonumber
\end{align}
which add to the corresponding one-loop expressions.
Again, we have introduced the unspecified coefficients $a_i,b_i,c_i \in\mathbbm{R}$ and appropriate factors of $N$.
With these structural two-loop expressions, the behavior of the order $\epsilon^2$ term of the GNY fixed point values for $y^2$ and $\lambda$ is determined to scale as $y^{2}_\ast(\epsilon^2)\sim\epsilon^2/N^2$ and $\lambda_\ast(\epsilon^2)\sim \epsilon^2/N^2$ for large $N$ which is consistent with Ref.~\onlinecite{Rosenstein:1993zf}.
Therefore, together with the one-loop contributions to $y^2_\ast$ and $\lambda_\ast$, we find that the derivative of Eq.~\eqref{eq:fast2l} w.r.t. $g^2$ scales as $1/N$ and does not provide a non-trivial contribution to $\theta$ in the limit $N\to \infty$.

The only potentially non-trivial term that is left to discuss at order $\epsilon^2$ is the contribution from the term $\sim~\hspace{-0.1cm}3N y^2_\ast$ in Eq.~\eqref{eq:beta01}.
This contribution, however, is also suppressed at order $\epsilon^2$ due to the scaling $y^{2}_\ast(\epsilon^2)\sim\epsilon^2/N^2$, i.e.
\begin{align}
	\theta^{\text{(2-loop)}}(N\to\infty)=\theta^{\text{(1-loop)}}(N\to\infty)=2-2\epsilon\,.
\end{align}
We conjecture this argument to perpetuate to higher loop orders as the introduction of any closed fermion loop coming with a factor $N$ comes with two Yukawa vertices and introduces another factor of $y^{2}_\ast\sim 1/N$ at the fixed point, suggesting $\theta_2^{\text{(exact)}}(N\to\infty)=2-2\epsilon$.

\subsection{Gauge field in $(4-\epsilon)$ dimensions}\label{sec:charge}

In this section, we study the case where the gauge field propagates in the same number of dimensions as the fermions and complex boson fields.
Then, the $\beta$ function for the charge is given by Eq.~\eqref{eq:betae} and charge parameter $e^2$ features a non-trivial fixed point,
\begin{align}\label{eq:east}
e^{2}_\ast =\frac{3}{4 N}\epsilon\,.
\end{align}
Here, the charge corresponds to an RG irrelevant direction, as $(\partial\beta_{e^2}/\partial e^2)|_{e^2=e^{\ast 2}}=-\epsilon$.
We deduce the fixed-point value of the Yukawa coupling, Eqs.~\eqref{eq:betay} and \eqref{eq:east},
\begin{align}\label{eq:fpymode}
	y^{2}_\ast =\frac{1+6 (e^{2}_\ast/\epsilon)}{N+1}\epsilon=\left(1+\frac{9}{2N}\right)\frac{\epsilon}{N+1}\,,
\end{align}
which then allows to calculate the fixed-point value of the quartic coupling
\begin{align}\label{eq:lambdaast}
	\lambda_\ast=\frac{C_N-8-N}{10(1+N)}\epsilon\,,
\end{align}
where $C_N=\sqrt{N^2+56N+424+810/N}$ and $\lambda_\ast>0$ for all $N$.
For large $N$, $\lambda_\ast$ vanishes like $\lambda_\ast\sim 2\epsilon/N$.
The RG scaling related to the interaction parameters $y^2$ and $\lambda$ can be derived from the stability matrix block, Eq.~\eqref{eq:stabmat}, where $\bar e^2$ is replaced by $e_\ast^2$.
This provides the eigenvalues $m_1=-\epsilon(1+9/(2N))$ and $m_2=-\epsilon\, C_N/(1+N)$ which are negative for all $N$.

\begin{figure}[t!]
\includegraphics[width=.9\columnwidth]{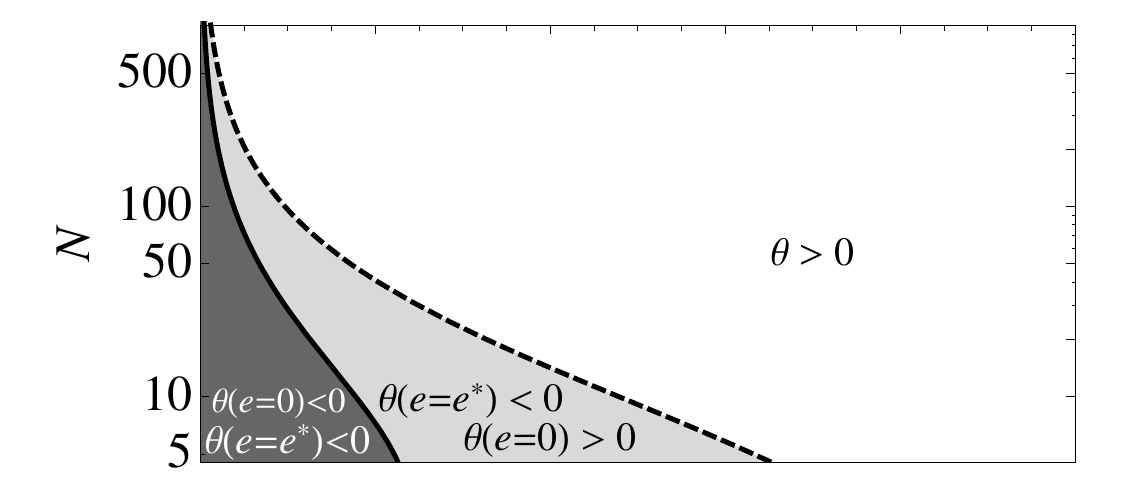}
\includegraphics[width=.9\columnwidth]{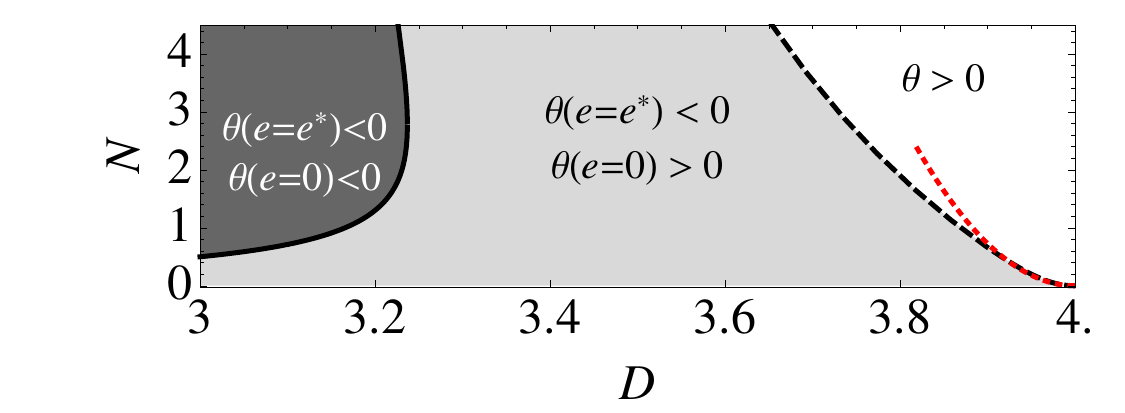}
\caption{Sign of the critical exponent resulting from the presence of the cubic coupling as a function of the dimensionality $D$ and the number of fermion flavors $N_f$ in first-order $\epsilon$ expansion.
The gray-shaded areas show the regions, where $\theta_2<0$. The two shadings distinguish the situations, where the coupling to the gauge field vanishes $e=0$ (darker gray) and where it is at its own fixed point $e=e^\ast$ (lighter gray). Red red dotted line shows Eq.~\eqref{eq:nfcrit}.}
\label{fig:theta2}
\end{figure}

Accordingly, the RG scaling of the cubic term is given by
\begin{align}\label{eq:theta2e2}
	\theta=\frac{\partial \beta_{g^2}}{\partial g^2}\Big|_{e_\ast^2,y^2_\ast,\lambda_\ast,g_\ast^2}=2-\frac{77+14 N +6C_N}{10 (1+N)}\epsilon\,.
\end{align}
In Fig.~\ref{fig:theta2}, the region where $\theta$ becomes negative in this scenario, is indicated by the shaded region which is separated by the dashed line from the region where $\theta>0$.
As two interesting numerical examples, we explicitly note that $\theta(N=2,\epsilon)$ becomes negative for $\epsilon\gtrsim 0.21$ or $D \approx 3.79$ and $\theta(N=1,\epsilon)$ for $\epsilon\approx 0.13$ or $D=3.87$.

In the limit of large $N$, we again find $\theta(N\to\infty)\to 2-2\epsilon$, so a sign change of $\theta$ requires a rather large value for $\epsilon$.
In contrast, for small $N$ we can simplify Eq.~\eqref{eq:theta2e2} and calculate the critical number $N_c$ for a given $\epsilon$ where $\theta$ becomes negative,
\begin{align}\label{eq:nfcrit}
	\theta\approx 2-\frac{27}{5}\sqrt{\frac{10}{N}}\epsilon=0\Rightarrow N_{c}\approx\frac{729}{10}\epsilon^2\,.
\end{align}
For small $N$ the position where $\theta$ changes sign, converges to  $N\to 0$ as $D$ approaches 4, see the red dotted line in Fig.~\ref{fig:theta2}.
This limit, however, has to be considered with some care, as for a given $\epsilon$ the fixed point values can grow large when $N$ decreases.
Therefore, the position where the sign change appears, e.g., in terms of dimensionality $D$, can be associated with large interaction parameters.
On the other hand, for all physically relevant $N$, i.e. $N\in \{1/2,1,2,...\}$ the fixed point values of all couplings remain of order $\mathcal{O}(1)$ in units of $\epsilon$.

\section{Summary \& Conclusions}\label{sec:conc}

We studied the quantum phase transition of $U(1)$ charged Dirac fermions coupled to a Kekul\'e order parameter with $Z_3$ symmetry.
In $4-\epsilon$ dimensions, the canonically relevant cubic term, which is allowed by the symmetry, can be rendered irrelevant through quantum fluctuations. When the charge coupling vanishes this occurs only for large values of $\epsilon$. A finite charge coupling, however, can support strongly the scenario where the renormalization group direction corresponding to the cubic coupling is rendered irrelevant.
In this case, the system gives rise to quantum critical behavior. We have investigated two scenarios:

(1) The charge is an exactly marginal coupling, as the gauge field propagates in four spacetime dimensions.
In that case, the value of the charge parameter $\bar e^2$ may be chosen freely, to control the position of the boundary where the critical exponent, determining the relevance of the cubic term, changes sign. Small values of the charge, $\bar e^2 \lesssim 1/10$, move this boundary from being close to three dimensions to  being very close to four (space-time) dimensions, bringing it under perturbative control.

(2) The charge coupling features a non-trivial fixed point value, when the  gauge field propagates in the same $4-\epsilon$ dimensions as the matter fields.
Here, the presence of a finite charge fixed-point coupling also strongly supports the emerging irrelevance of the cubic term at smaller values of $\epsilon$.
For a small number of fermion flavors, the boundary where the coupling is rendered irrelevant moves close to four dimensions, however, at the cost of the prefactors of the $\sim \epsilon$ fixed point values growing large.

Furthermore, we have carefully discussed the limit of a large number of Dirac fermion flavors in the case of vanishing charge, as well as for finite charge in the scenario~(2). In this limit, the cubic coupling always provides a renormalization group relevant direction unless $\epsilon>1$,  which is again beyond the domain of validity of this approach. Finally, for vanishing charge we have shown that this statement also holds at two-loop order, and provided an argument that suggests that this conclusion continues to hold at arbitrary order in the loop expansion.

\begin{acknowledgments}
The authors are grateful to Fakher Assaad and Igor Boettcher for useful discussions. M.M.S. is supported by DFG Grant No. SCHE 1855/1-1 and I. F. H. is supported by the NSERC of Canada.
\end{acknowledgments}




\end{document}